\documentclass{article}

\usepackage{arxiv}

\usepackage[utf8]{inputenc} 
\usepackage[T1]{fontenc}    
\usepackage{hyperref}       
\usepackage{url}            
\usepackage{booktabs}       
\usepackage{amsfonts}       
\usepackage{nicefrac}       
\usepackage{microtype}      
\usepackage{lipsum}

\usepackage{amsthm,amsmath,amssymb}
\usepackage{graphicx}
\usepackage{subcaption}
\usepackage{listings}
\usepackage{courier}
\usepackage[dvipsnames]{xcolor}
\usepackage[justification=centering,skip=0pt,labelfont=bf]{caption}

\usepackage{etoolbox}
\newtoggle{InString}{}
\togglefalse{InString}
\newcommand*{\ColorIfNotInString}[1]{\iftoggle{InString}{#1}{\color{blue}#1}}%
\newcommand*{\ProcessQuote}[1]{#1\iftoggle{InString}{\global\togglefalse{InString}}{\global\toggletrue{InString}}}%
\newenvironment{figureAsListing}
    {
    \addtocounter{figure}{-1}
    \refstepcounter{lstlisting}

    \begin{figure}
        \vspace{-1em}
        \centering
    }
    { 
        \end{figure} 
    }
\def\O{\mathcal{O}}  

\title{In Search of the Fastest \\ Concurrent Union-Find Algorithm}

\author{
  Dan Alistarh\\
  IST Austria\\
  Klosterneuburg, Austria\\
  \texttt{dan.alistarh@ist.ac.at} \\
  \And
  Alexander Fedorov \\
  Higher School of Economics and JetBrains\\
  St. Petersburg, Russia \\
  \texttt{afedorov2602@gmail.com} \\
  \And
  Nikita Koval \\
  IST Austria and JetBrains\\
  Klosterneuburg, Austria and St. Petersburg, Russia\\
  \texttt{ndkoval@ya.ru}
}

\begin{document}

\maketitle

\begin{abstract}
Union-Find (or Disjoint-Set Union) is one of the fundamental problems in computer science; it has been well-studied from both theoretical and practical perspectives in the sequential case. 
Recently, there has been mounting interest in analyzing this problem in the concurrent scenario, and several asymptotically-efficient algorithms have been proposed. Yet, to date, there is very little known about the practical performance of concurrent Union-Find.

This work addresses this gap. We evaluate and analyze the performance of several concurrent Union-Find algorithms and optimization strategies across a wide range of platforms (Intel, AMD, and ARM) and workloads (social, random, and road networks, as well as integrations into more complex algorithms).  
We first observe that, due to the limited computational cost, the number of induced cache misses is the critical determining factor for the performance of existing algorithms. We introduce new techniques to reduce this cost by storing node priorities implicitly and by using plain reads and writes in a way that does not affect the correctness of the algorithms. 
Finally, we show that Union-Find implementations are an interesting application for  Transactional Memory (TM): one of the fastest algorithm variants we discovered is a sequential one that uses coarse-grained locking with the lock elision optimization to reduce synchronization cost and increase scalability. 
\end{abstract}



\section{Introduction}
\vspace{-1em}

\emph{Union-Find} {---} the problem of maintaining the global connectivity structure of a set of integer elements based on their pair-wise connectivity {---} is a fundamental problem in computer science. 
This problem is alternatively known as Disjoint-Set Union (DSU), and is a key component of several classic algorithms, such as Kruskal's \cite{kruskal} and Boruvka's \cite{boruvka} algorithms for finding minimum spanning trees (MSTs), maintaining connected components and finding loops under edge additions, or finding strongly-connected components in directed graphs. 

In its classic formulation, the sequential DSU problem assumes a finite ground set $S$, upon whose elements we perform the following operations: 
\vspace{-0.5em}
\begin{itemize}
    \item \texttt{SameSet(u, v)} checks whether two elements \texttt{u} and \texttt{v} are in the same set;
    \item \texttt{Union(u, v)} merges the sets to which \texttt{u} and \texttt{v} are currently assigned;
    \item \texttt{Find(u)} returns the representative of the set in which \texttt{u} is located. This representative must be the same for all elements from the corresponding subset. 
\end{itemize}

It is worth noting that in sequential case \texttt{SameSet} can be easily implemented via two \texttt{Find} invocations, on \texttt{u} and \texttt{v}, and checking whether the representatives coincide. Further optimizations exist to implement this method more efficiently; we discuss them in Section~\ref{section:optimizations}. 

This classic problem is known to have a rich theoretical structure. A masterclass by Tarjan linked the upper bound on the worst-case time complexity of the problem to the inverse Ackermann function~\cite{Tarjan1975}, followed by a matching lower bound for a restricted case~\cite{tarjan1979class}, which was later extended to arbitrary algorithms to show optimality~\cite{Fredman1989}.
 Later, Tarjan and Van Leeuwen~\cite{Tarjan1984} performed one of the first worst-case analyses for compaction heuristics in the context of Disjoint-Set Union (DSU), exposing the highly non-trivial fact that, with appropriate path compaction and linking heuristics, the problem can be solved in $\O(m \cdot \alpha(n, \frac m n))$ time complexity, where $n$ is the number of elements, $m$ is the number of operations, and $\alpha$ is a functional inverse of Ackermann's function. 

Patwary, Blair, and Manne~\cite{Patwari} were the first to perform an in-depth experimental study of sequential strategies. Their work clearly showed the importance of minimizing the number of reads from memory on the performance of DSU, since the various strategies have very limited computational demands. Their experiments exposed the fact that the fastest sequential algorithm was the one designed by Rem in 1976~\cite{Rem}, which we describe in detail in the following sections.

The first concurrent algorithm for DSU was proposed by Cybenko, Allen, and Polito~\cite{Cybenko1988}, whose key idea was using a spin-lock for write operations. Years later, Anderson and Woll proposed a \emph{wait-free} concurrent algorithm \cite{Anderson94}, which is roughly a concurrent generalization of one of the linking-and-compression strategies studied by Tarjan and Van Leeuwen~\cite{Tarjan1984}.
Their paper claims a worst-case upper bound for the algorithm of $\Theta(m \cdot (\alpha(m, 0) + p))$, where $p$ is the number of parallel processes, and $m$ is the number of operations; it was later observed that this proof is correct only under the non-standard assumption that threads cannot be preempted in between some certain operations in the algorithm~\cite{Jayanti2016}.

Recently, Jayanti and Tarjan presented a set of correct and asymptotically-efficient concurrent DSU algorithms, which use fixed random priorities rather than ranks, based on the randomized sequential algorithm of \cite{Tarjan2014}. These algorithms achieve a (total) work complexity upper bound of $\mathcal{O}(m \cdot (\alpha(n, \frac {m} {p}) + \log (\frac {np} m + 1)))$.
Recent work by these authors, in collaboration with Boix-Adserà, suggested an algorithm with the same total work complexity bounds, but showed that this is optimal for a class of natural ``symmetric algorithms''~\cite{Jayanti2019}.

\paragraph*{Our Contribution.}
Motivated by the significant recent interest in the concurrent DSU problem, as well as by its numerous practical applications, in this paper we perform the first thorough study of the practical performance of concurrent DSU implementations. 
While our focus is to implement and study existing algorithmic proposals, along the way we discover new optimizations and algorithmic insights. 
We start from the basic observation that, given the simple structure of most algorithms for DSU, memory access and synchronization costs will be the dominating factors behind practical performance. 
With this in mind, we analyze the performance of several classic baselines, and propose a host of optimizations to specifically reduce the impact of these factors. 

We perform a range of experiments across several architectures, algorithm variants and optimizations, synchronization primitives, as well as workloads, to determine the fastest concurrent DSU algorithm. We provide a wide range of results and discuss our findings across several dimensions in detail in Section~\ref{sec:discussion}. 
In brief, the variant which appears to be ``fastest'' for most of the settings considered is an optimized sequential algorithm variant that leverages HTM (lock elision) for high path compaction while minimizing synchronization cost. 

\section{Experimental Setup}
\vspace{-1em}

For our evaluation, we use two classic graph algorithms based on DSU data structures. The first  maintains the \emph{connected components} for a given graph. In this case, for benchmarking, we randomly split graph edges between threads, and also mark whether the \texttt{Union} of the \texttt{SameSet} operation should be executed with a given edge as a parameter. The set of graphs we consider is presented in Table 1. 

The second benchmark is a parallel version of the \emph{Boruvka's algorithm} \cite{boruvka} for finding the minimum spanning tree. (The pseudo-code of the algorithm we use in the experiments is presented in Listing~\ref{lst:boruvka}.) This algorithm performs at most $\log n$ iterations (where $n$ is number of vertices), each is split into two phases. During the first phase, the algorithm finds the shortest adjacent edge for each vertex, removing the ones that connect vertices from the same component, the set of which is maintained by the DSU data structure. Here, the \texttt{UpdateIfShorter} function atomically checks whether the already stored edge is longer than the specified one, and replaces it in this affirmative case. During the second phase, the algorithm goes through all the representatives of the components and adds the corresponding shortest edges to the MST. The main idea of the concurrent algorithm is that both phases can be performed in parallel, with synchronization between them. However, if the number of remaining components is small, it is better to complete the work sequentially. Thus, we perform the parallel part until the already built MST part exceeds some size threshold. We consider only the parallel part by the benchmark. 

\begin{figureAsListing}
\begin{lstlisting}
func ParallelMST(G: Graph) {
  mst = {} // empty set initially    
  while mst.Size < THRESHOLD(G.Size) {
    shortestEdges: Edge[G.Size] // array of the shortest edges for each component
    // Phase 1. Find the shortest edges for each vertex
    parallel for e in G.Edges {
      if SameSet(e.from, e.to) { G.Edges.remove(e); continue }
      UpdateIfShorter(&shortestEdges[Find(e.from)], e)
      UpdateIfShorter(&shortestEdges[Find(e.to)], e)
    }
    // Phase 2. Unite components and update MST
    parallel for u in G.Vertices {
      Unite(shortestEdges[u].from, shortestEdges[u].to)
      mst += shortestEdges[u]
    }
  }
  return mst + SequentialMST(graph)
}
\end{lstlisting}
\caption{Parallel Boruvka's algorithm pseudo-code used in our experiments}
\label{lst:boruvka}
\end{figureAsListing}

\paragraph*{Graph Inputs.}
We evaluate the algorithms above on a range of real-world and synthetic graphs. The connected components maintenance algorithm has linear time complexity (assuming for simplicity that the DSU operations work in constant time), while  Boruvka's algorithm requires logarithmic time to complete, which is  why we test it on relatively smaller graphs. Nevertheless, both benchmarks use road and social network graphs, as well as random graphs with similar properties. The list of graphs we use for the connected components and Boruvka's algorithm benchmarks is presented in Tables~\ref{tabel:connected_components_graphs}~and~\ref{table:boruvka_graphs} respectively. 

The first graphs in both tables represent the USA road network, Central and West parts. The next two graphs for the connected components benchmark are social network graphs, while the ones for the Boruvka's benchmark are Internet topology graphs. Also, we use two large synthetic graphs. The first one is a randomly-generated Erdoes-Renyi graph, with a specified number of vertices and edges ($2.5M$ vertices and $30M$ edges for the connected components, and $1M$ vertices and $10M$ edges for the Boruvka's algorithm); it is denoted as \texttt{RANDOM} in our experiments. The second synthetic graph is folklore for detecting ``bad'' DSU implementations in programming contests---most of the edges are incident to a small number of nodes; it is denoted as \texttt{HIGH-CONTENTION}.

\begin{table}
    \centering
    \begin{tabular}{|p{3.6cm}|p{2.05cm}|p{1.75cm}|p{4.7cm}|}
    \hline
\large{Graph} & \large{Vertices (M)}&\large{Edges (M)} & \large{Description}    \vspace{0.07cm}
\\
\hline
 \textbf{USA-ROADS} & \centering\large14.1  & \centering\large16.9  & A graph with roads of the central part of the USA \\\hline
  \textbf{LIVE-JOURNAL} & \centering\large4  & \centering\large34.7  & A graph of ground-truth communities in LiveJournal social network \\\hline
 \textbf{POKEC} & \centering\large1.6  & \centering\large30.5  & A graph of friendship relations in Pokec social network \\\hline
\textbf{RANDOM} & \centering\large2.5  & \centering\large30  & A random graph generated with the Erd\H{o}s-R\'enyi model. \\\hline
 \textbf{HIGH-CONTENTION} & \centering\large12  & \centering\large26.8  & A synthetic graph for detecting ``bad'' DSU versions\\
 \hline
    \end{tabular}
    \vspace{0.7em}
    \caption{Graphs for the connected components benchmark.}
    \label{tabel:connected_components_graphs}
\end{table}

\begin{table}
    \centering
    \begin{tabular}{|p{3.95cm}|p{2.05cm}|p{1.75cm}|p{4.4cm}|}
    \hline
\large{Graph} & \large{Vertices (M)}&\large{Edges (M)} & \large{Description} \vspace{0.07cm}\\
\hline
 \textbf{USA-ROADS} & \centering\large6.2  & \centering\large7.6  & A graph with roads of the western part of the USA \\\hline
 \textbf{BERKELEY-STANFORD} & \centering\large6.9  & \centering\large7.6  & Hyperlinks between the Berkeley and the Stanford domains \\\hline
 \textbf{INTERNET-TOPOLOGY} & \centering\large1.7  & \centering\large11.1  & Internet topology graph, from traceroutes run daily in 2005\\\hline
 \textbf{RANDOM} & \centering\large1  & \centering\large10  & A random graph generated with the Erd\H{o}s-R\'enyi model \\\hline
 \textbf{HIGH-CONTENTION} & \centering\large4  & \centering\large8.9  & A synthetic graph for detecting ``bad'' DSU versions \\ 
 \hline
    \end{tabular}
    \vspace{0.7em}
    \caption{Graphs for the Boruvka's algorithm benchmark}
    \label{table:boruvka_graphs}
\end{table}

\paragraph*{Hardware.}
We perform the experiments on Intel, AMD, and ARM platforms; all the machines we used have several sockets, which induces Non-Uniform Memory Access (NUMA) effects. The detailed specifications are as follows:
\begin{itemize}
    \item \textbf{Intel}. We used our local machine with 4 sockets, Intel Xeon Gold 6150 with 18 cores per socket, and hyperthreading enabled, for 144 hardware threads in total.
    \item \textbf{AMD}. We used a general-purpose Amazon AWS \cite{amazon-aws} instance with 6 sockets, 8-core AMD EPYC 7571 processors with hyperthreading enabled in each; 96 cores total. 
    \item \textbf{ARM}. We used an instance on the Packet \cite{packet} cloud service with 2 sockets of 48-core ARM Cavium ThunderX processors; 96 cores total. The ARM memory model is more relaxed than the TSO one on Intel and AMD architectures, which would lead us to expect higher scalability.
\end{itemize}

\paragraph*{Software.}
All algorithms and benchmarks are implemented either in Java or Kotlin, and compiled to the JVM byte-code; we use OpenJDK 11.0.4 with Ubuntu OS on all platforms. To avoid problems related to JIT compilers and reproducibility/benchmarking, we use the Java Microbenchmark Harness (JMH) library to run our benchmarks and collect statistics \cite{jmh}.

\section{Sequential Implementations via Compressed Trees}
\vspace{-1em}
We begin with an overview of sequential implementations, which are loosely based on the idea of maintaining a \emph{compressed forest of trees} \cite{Tarjan1975}. 
Each tree in this data structure corresponds to the membership of one set, where the root of the tree acts as its representative. The trees are implemented by maintaining parents for each element; thus, the data structure stores an array of parent links. Roots of the trees have their parent links point to themselves. Thus, in order to implement the \texttt{Find} operation, the algorithm ``climbs'' using parent links until it reaches the root. At the same time, the \texttt{Union} operation takes the corresponding tree roots as representatives, checks whether they coincide (finishing the operation in that case), and unites the sets by pointing one root to another. While the algorithm seems straightforward, it requires some heuristics to guarantee good time complexity. 

\paragraph*{Linking Strategies.} 
When the \texttt{Union} operation decides to merge two different sets, it either changes the parent link of the first element to the second one or vice versa. Intuitively, we want to maintain the tree height as small as possible, so that \texttt{Find} operation works efficiently. The standard procedure is defining priorities on roots, so that the root with the lower priority is ``hung'' under the root with higher priority. The following definitions of priority are usually employed:
\vspace{-0.5em}
\begin{itemize}
    \item \emph{Tree size}. Each root maintains the size of its tree, and smaller roots are pointed to larger ones. This serves as a way of balancing the tree.
    \item \emph{Rank}. Since we aim at optimizing the height, it is reasonable to store the height as a priority. When the ranks of the trees to be united are different, then the smaller tree is pointed to the larger one, and the ranks remain the same. However, when the ranks coincide, an arbitrary one is chosen as a new root, and its rank is incremented.
    \item \emph{Random.} Another method to choose a new root is using a set of fixed random initial priorities; Tarjan et. al. first analyzed this technique in the sequential case~\cite{Tarjan2014}, and after it Jayanti and Tarjan made an analysis for the concurrent case \cite{Jayanti2016}. 
\end{itemize}

\paragraph*{Path Compaction.} 
Another way to make the data structure faster is making the tree ``flatter'' by shortening paths between nodes and the roots. In particular, notice that during the \texttt{Find} operation, the parent links can be changed to the higher ones without breaking correctness; this way, the algorithm reaches the root faster on the next \texttt{Find} invocation. Here are several strategies which are usually employed to compact trees:
\begin{itemize}
    \item \emph{Compression.} Once the root is found, all the elements on the search path can update their parent pointers to the root. The simplest implementation uses recursion and performs these updates in the last-in-first-out order. The disadvantage of this technique is that it requires performing two passes, from $u$ to the root and back to $u$; thus, in practice, this strategy produces extra cache misses.
    \item \emph{Splitting.} This technique updates the parent links during the only traversal to the root. On each step, it reads both the parent and the grandparent of the current element, and updates its parent link to the grandparent. Thus, it compresses the paths from all visited elements to the root by one.
    \item \emph{Halving.} This technique is similar to splitting, but after each step, each node is linked to the grandparent. Thus, parents of only half of the elements on the path are updated.
\end{itemize}

These compaction optimizations can guarantee $\O(\log n)$ amortized time complexity of the \texttt{Find} operation, while the linking techniques guarantee $\O(\log n)$ worst-case time complexity (probabalistic in case of random priorities). However, using both techniques at the same time, linking by priority and path compaction, guarantees $\O(m \cdot \alpha(n, \frac m n))$ time complexity on average, where $n$ is the number of elements, $m$ is the number of operations, and $\alpha$ is a functional inverse of Ackermann's function. 

\paragraph*{Implementation Example.}
Listing~\ref{lst:sequential} provides a pseudo-code of the DSU algorithm with ranks as priorities and path compaction via halving. Here we assume that the \texttt{SameSet} function is implemented with two \texttt{Find} invocations.

\begin{figureAsListing}
\begin{minipage}[t]{0.49\textwidth}
\begin{lstlisting} 
func Union(u, v) {
  u = Find(u); v = Find(v)
  if u == v: return
  if rank[u] <= rank[v]:
    parent[u] = v
  else if rank[u] > rank[v]:
    parent[v] = u
  if rank[u] == rank[v]: rank[v]++
}
\end{lstlisting}
\end{minipage}
\hfill
\begin{minipage}[t]{0.49\textwidth}
\begin{lstlisting}[firstnumber=10]
func Find(u) = while (true) {
  p = parent[u]
  gp = parent[p] 
  if p == gp: return p
  parent[u] = gp
  u = gp
}
\end{lstlisting}
\end{minipage}
\caption{Sequential DSU with rank-based priorities using path compaction via halving.}
\label{lst:sequential}
\end{figureAsListing}

\paragraph*{Evaluation.}

\begin{figure}
\centering
\includegraphics[width=0.7\textwidth]{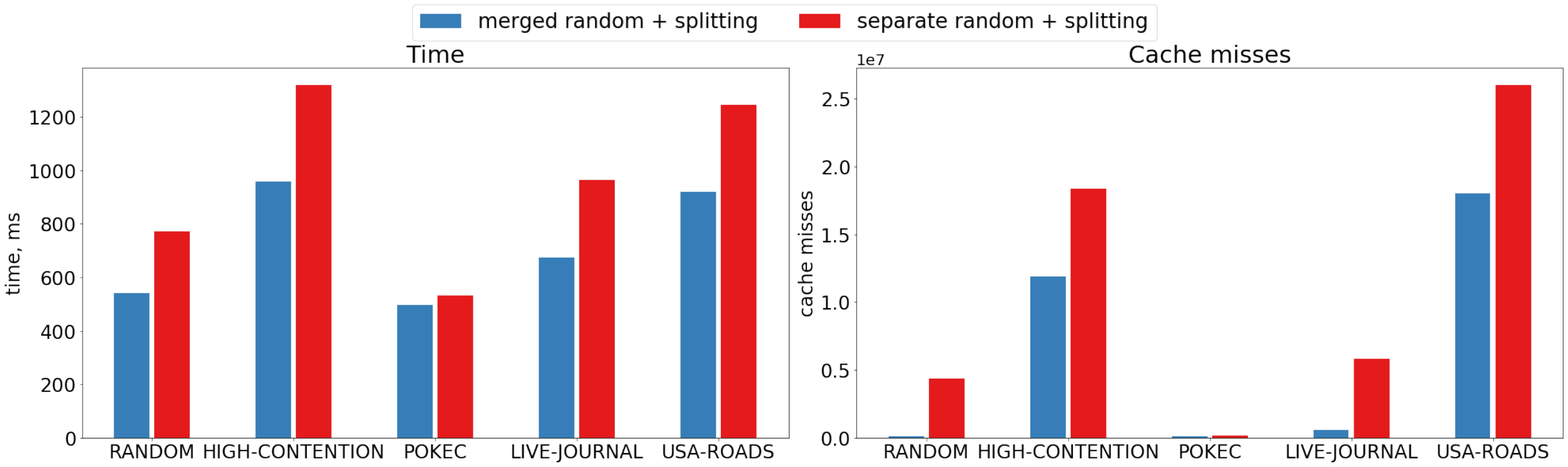}
\vspace{-5em}
\caption{Here we consider two sequential DSU algorithm versions evaluated on the connected components benchmark, which use different ways to store parents and priorities. While both versions use splitting path compaction strategy and random priorities, the red one stores parent links and priorities in separate fields, whereas the blue one combines them into one field by stealing one bit to define whether the node is a root (stores priority) or not (stores parent). Since the first variant uses twice less memory, it induces much fewer cache misses and significantly improves the performance. Moreover, some graphs (with ~4 million nodes or less) are entirely stored in the cache, so that there are almost no cache misses in their evaluation.\vspace{2em}}
\label{fig:cache-misses}
\end{figure}

We evaluated all the combinations of linking and path compaction strategies. Similarly to the work by Patwary et al., we store priorities only for roots \cite{Patwari}. Thus, it is possible to use a single register to store either parent or priority {---} we only have to reserve one bit as a mark whether the element is a root or not; Figure~\ref{fig:cache-misses} shows that this optimization significantly improves the performance by reducing the number of cache misses.

In addition to the standard linking strategies, we suggest using a pseudo-random one, which essentially shuffles priorities in the range $1..n$ among $n$ elements {---} the following well-known hashing formula due to Knuth is used:
\begin{equation*}
priority(x) = (x + \mathtt{SHIFT}) \times \mathtt{BIG\_PRIME} \mod N.
\end{equation*}
It is worth noting that $\mathtt{SHIFT}$ can be chosen randomly to increase the algorithm's robustness against some adversary to make the algorithm work slow.

\begin{figure}
\centering
\includegraphics[width=\textwidth]{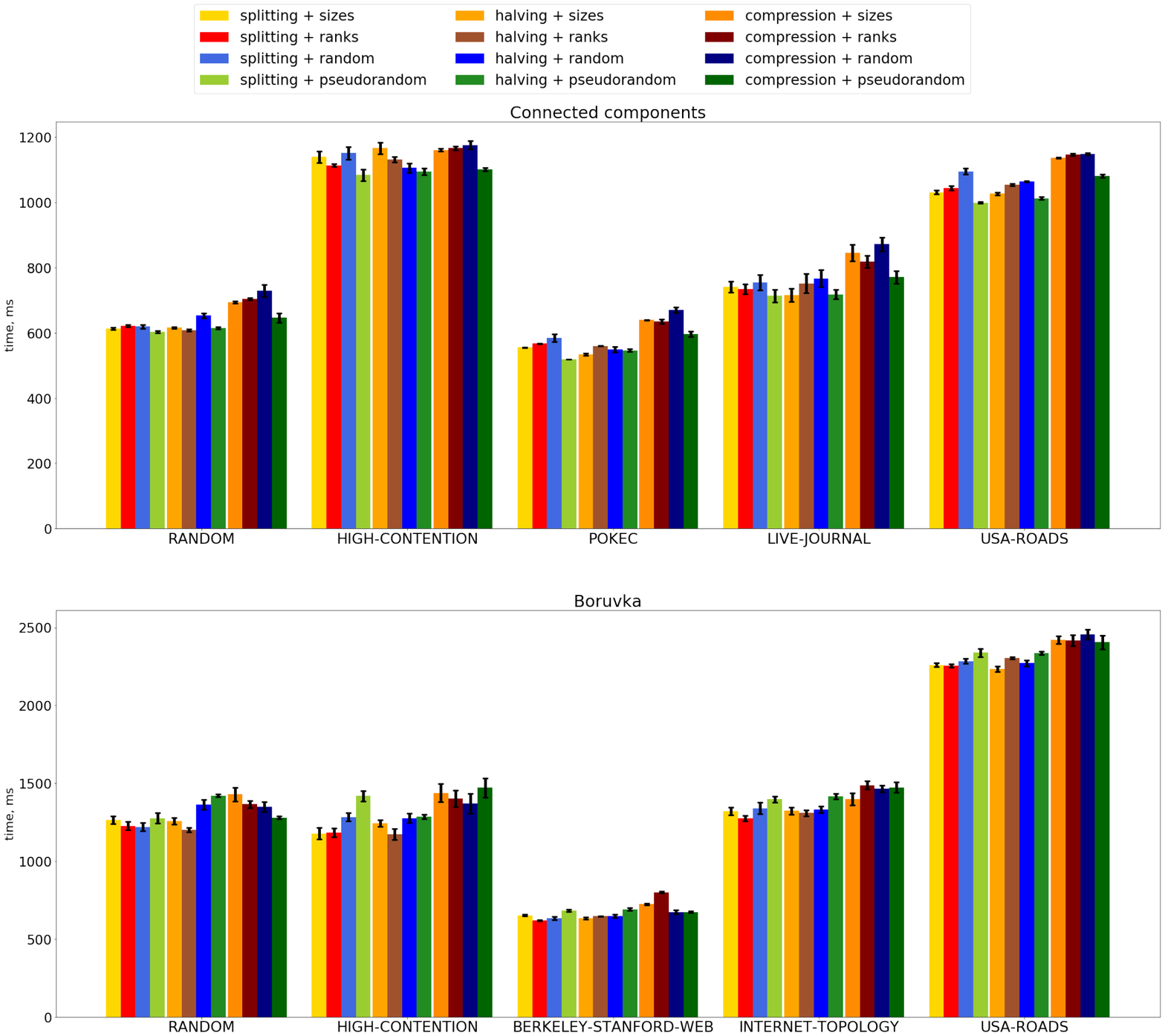}
\vspace{0.5em}
\caption{Time comparison for sequential DSU versions with different linking and path compaction strategies on the Intel machine on various input graphs. Lower is better.}
\label{fig:sequential}
\end{figure}

Figure~\ref{fig:sequential} shows the results of all the 12 combinations on both connected components and Boruvka's algorithm benchmarks; we used our local Intel machine for this experiment.

Rank priorities are approximations for tree sizes. We predictably do not see significant difference between these linking strategies; however, using ranks requires updates only if two trees in the \texttt{Unite} operation have the same ranks. Thus, we do not consider the size-based strategy in further concurrent tests {---} we believe that it cannot outperform the rank one due to more memory updates.

The suggested pseudo-random priorities slightly outperform the real random ones in almost all scenarios. We believe that this is a consequence of the fact that we do not need to store pseudo-random priorities at all and that the pseudo-random method shuffles elements as evenly as possible. It should be stated that in the experiment we do not include times of DSU initialization in evaluation times, so the reason for such results can not be more complex real random generation. Thus, we use only pseudo-random priorities in further experiments. As for the comparison between pseudo-random and rank-based priorities, we do not see one clearly and consistently outperform the other in many scenarios. Therefore, we consider it important to test both these strategies in a concurrent environment. Similarly, we keep all linking strategies for further experimentation.

\section{Evaluating Concurrent Implementations}

\subsection{Basic Variants}
In accordance with the evaluation of the sequential algorithms, we consider only rank-based and pseudo-random linking strategies combined with all three path compression techniques. Similarly to the existing concurrent DSU implementations, we achieve atomicity by merging trees via \texttt{Compare-And-Set (CAS)} primitive. 

Listing~\ref{lst:concurrent-impl} contains a pseudo-code for the concurrent version with ranks. For strategies with ranks we need to be sure that we for a not a rank and a parent can not be updated concurrently, otherwise we can get a non-linearizable behavior, where, for instance, two nodes have parent links pointing to each other violating the forest structure invariant. We can create a structure storing both a parent and a rank and change with \emph{Compare-And-Set} primitive the pointer to a structure, which was proposed by Anderson and Woll \cite{Anderson94}, but a faster way would be to store both a rank and a parent in the same register by either dividing it into two parts or by using the same trick as we used in the sequential case, when we stored priorities only for roots. The last technique was chosen since it uses twice less memory. A heuristic that was used for an optimization in the sequential case becomes important for the correctness in the concurrent case. Anderson and Woll also proposed that in the \texttt{Union} operation we should first update a parent of the joined node and then update the rank of the new root if needed, despite the fact that it can worsen time complexity due to possible atomicity violations.

\begin{figureAsListing}
\begin{minipage}[t]{0.49\textwidth}
\begin{lstlisting} 
func Union(u, v) = while (true) {
  (u, ru) = Find(u)
  (v, rv) = Find(v)
  if u == v: return
  if ru < rv {
    if CAS(&A[u], u, v): return
  } else if ru > rv {
    if CAS(&A[v], v, u): return
  } else { // ru == rv
    if u < v && CAS(&A[u], u, v):
      CAS(&A[v], rv, rv+1); return
    if u > v && CAS(&A[v], v, u):
      CAS(&A[u], ru, ru+1); return
  }
}
\end{lstlisting}
\end{minipage}
\hfill
\begin{minipage}[t]{0.49\textwidth}
\begin{lstlisting}[firstnumber=16]
func Find(u): (root, rank) {
  p = A[u]
  if isRank(p): return (u, p)
  (root, rank) = Find(p)
  if p != root: 
    CAS(&A[u], p, root)
  return (root, rank)
}

func SameSet(u, v) = while (true) {
  (u, _) = Find(u); (v, _) = Find(v)
  if u == v: return true
  if isRank(A[u]): // still a root?
    return false
}
\end{lstlisting}
\end{minipage}
\caption{Concurrent DSU with priorities via ranks and the path compression heuristic}
\label{lst:concurrent-impl}
\end{figureAsListing}

\begin{figure}
	\centering
	\vspace{-5em}
	\includegraphics[width=\linewidth]{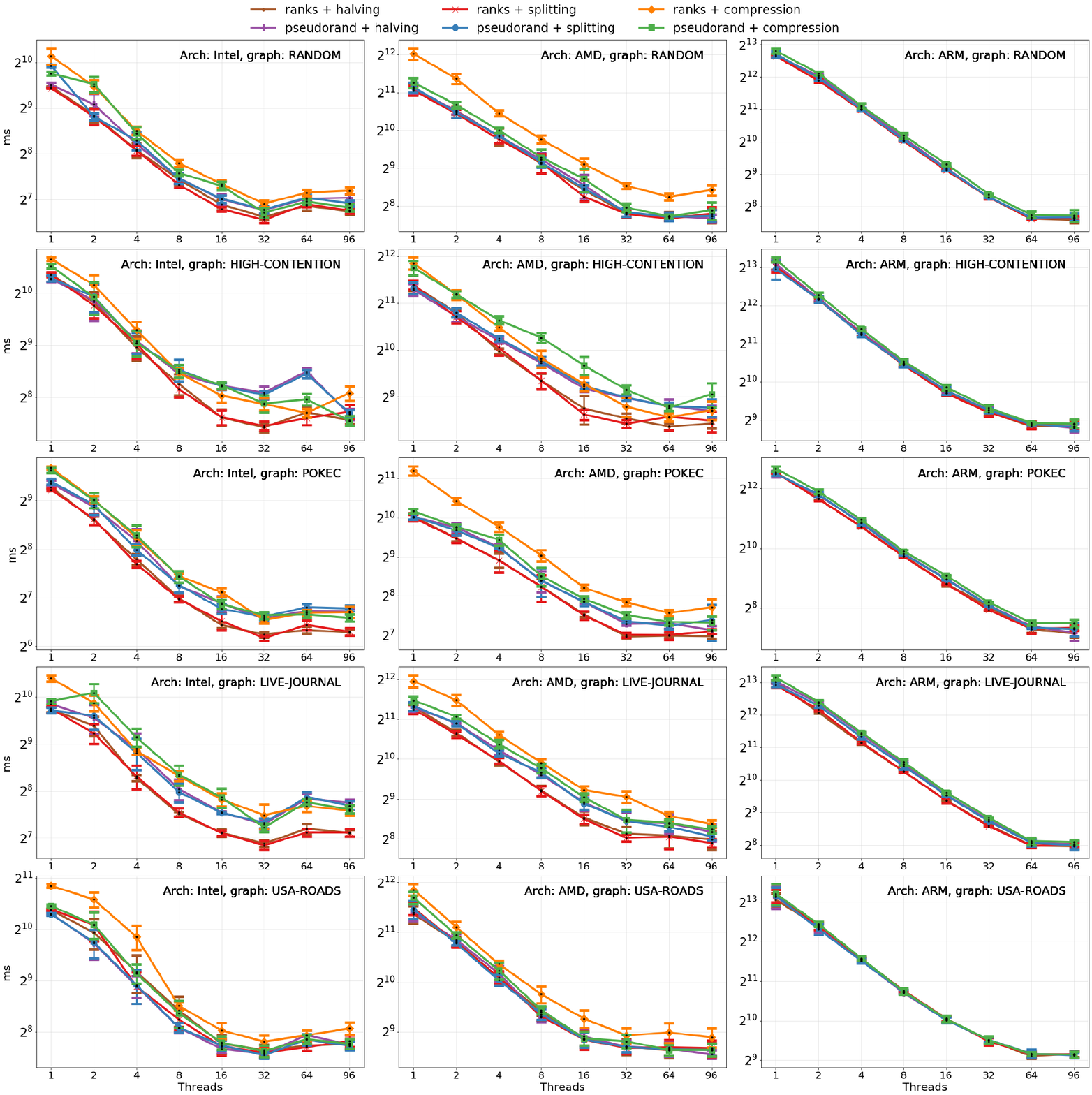}
\vspace{-9.5em}
\caption{Basic concurrent implementations comparison on the connected components benchmark. Each algorithm has been evaluated on five different graphs (see Table~\ref{tabel:connected_components_graphs}) and on Intel, AMD, and ARM platforms. Halving and splitting techniques show the best results, but the compression one is worse on both rank-based and pseudo-random strategies. While the trends on different architectures are similar, there are changes when NUMA effects appear, which can be noticed when threads begin working on different sockets. The versions on Intel and ARM machines do not scale after the moment they need to use two or more NUMA sockets, while the versions on AMD stop scaling only on three sockets. Another observation is that for ARM machines the difference between algorithms is less noticeable, because its processor is slower in comparison to other platforms, while the memory is as fast as in other machines, and we primarily optimize memory accesses.}
\label{fig:basic-concurrent}
\end{figure}

\begin{figure}
	\centering
	\vspace{-5em}
	\includegraphics[width=\linewidth]{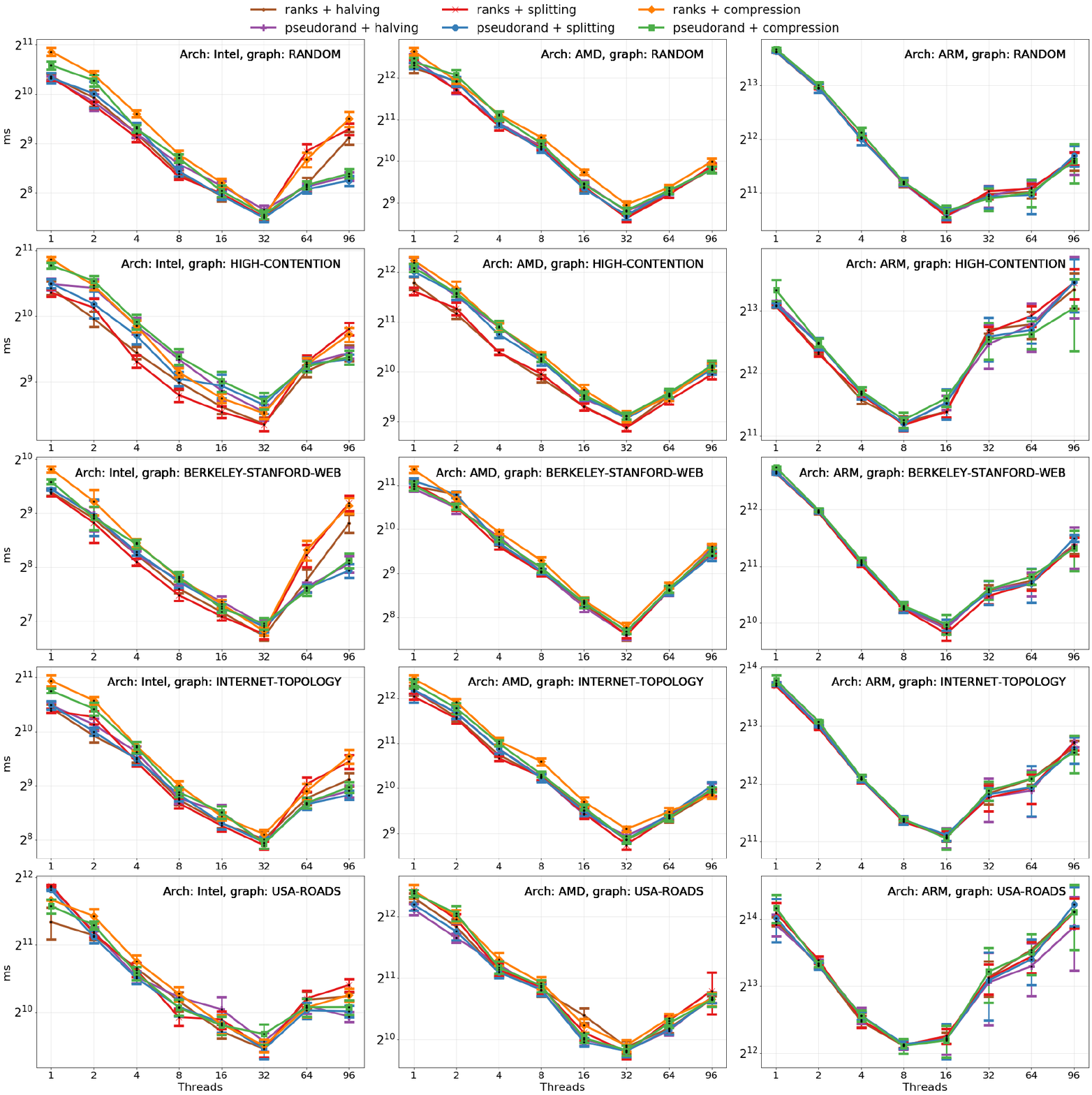}
	\vspace{-9.5em}
\caption{Basic concurrent implementations comparison on the Boruvka's algorithm benchmark. Here the algorithms do not scale well due to internal synchronization; thus, we are mainly interested in the fastest points {---} on 32 threads for Intel and AMD, on 8 or 16 threads for ARM.}
\label{fig:basic-boruvka}
\end{figure}

Figures~\ref{fig:basic-concurrent} and~\ref{fig:basic-boruvka} present the running time of the basic concurrent implementations for several of the strategies. 
One may expect that, following the sequential analysis, either rank-based or pseudo-random priorities would behave significantly better, but this is not apparent in the results, what motivates our further investigation. 
Another natural conjecture, which is apparent empirically, is that splitting and halving compaction strategies have very similar performance (since their cost is similar), while compression is inferior by comparison, due to two path traversals {---} for finding the root and for path compression. Thus, we use only splitting heuristic in our next experiments. We also have collected statistics with the numbers of cache misses, which shows that the algorithms with the compression technique get about $\times7$ more cache misses on loads and almost the same number on stores. 

\subsection{Optimizations}\label{section:optimizations}

\begin{figure}
	\centering
	\includegraphics[width=\linewidth]{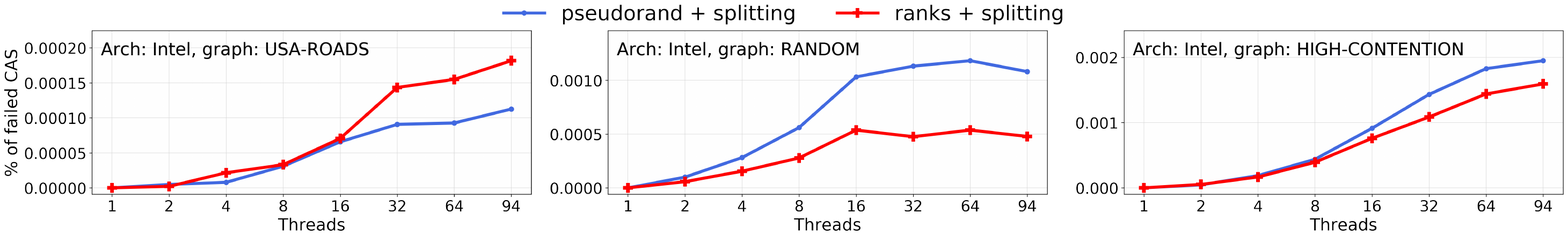}
\vspace{-12.5em}
\caption{The average number of failed \texttt{CAS} operations on the connected components benchmark (Intel platform). Here we see that almost all CAS operations succeed; thus, in practice, there is no reason to perform several attempts to update parents.}
\label{fig:failed-cas}
\end{figure}

\paragraph*{Path Compaction via Plain Writes.}
Figure~\ref{fig:failed-cas} shows the average number of failed CAS operations for different algorithms and thread numbers on the connected component benchmark; we used our local Intel machine to collect these statistics. In worst-case scenarios, only \texttt{0.002\%} of the total number of CAS invocations fail. Thus, it should be safe to consider that there is almost no contention in practice, and there is no reason to perform several attempts to update parents, which differs from the theoretical worst-case~\cite{Jayanti2016}.

\begin{figure}
	\centering
	\vspace{-5em}
	\includegraphics[width=\linewidth]{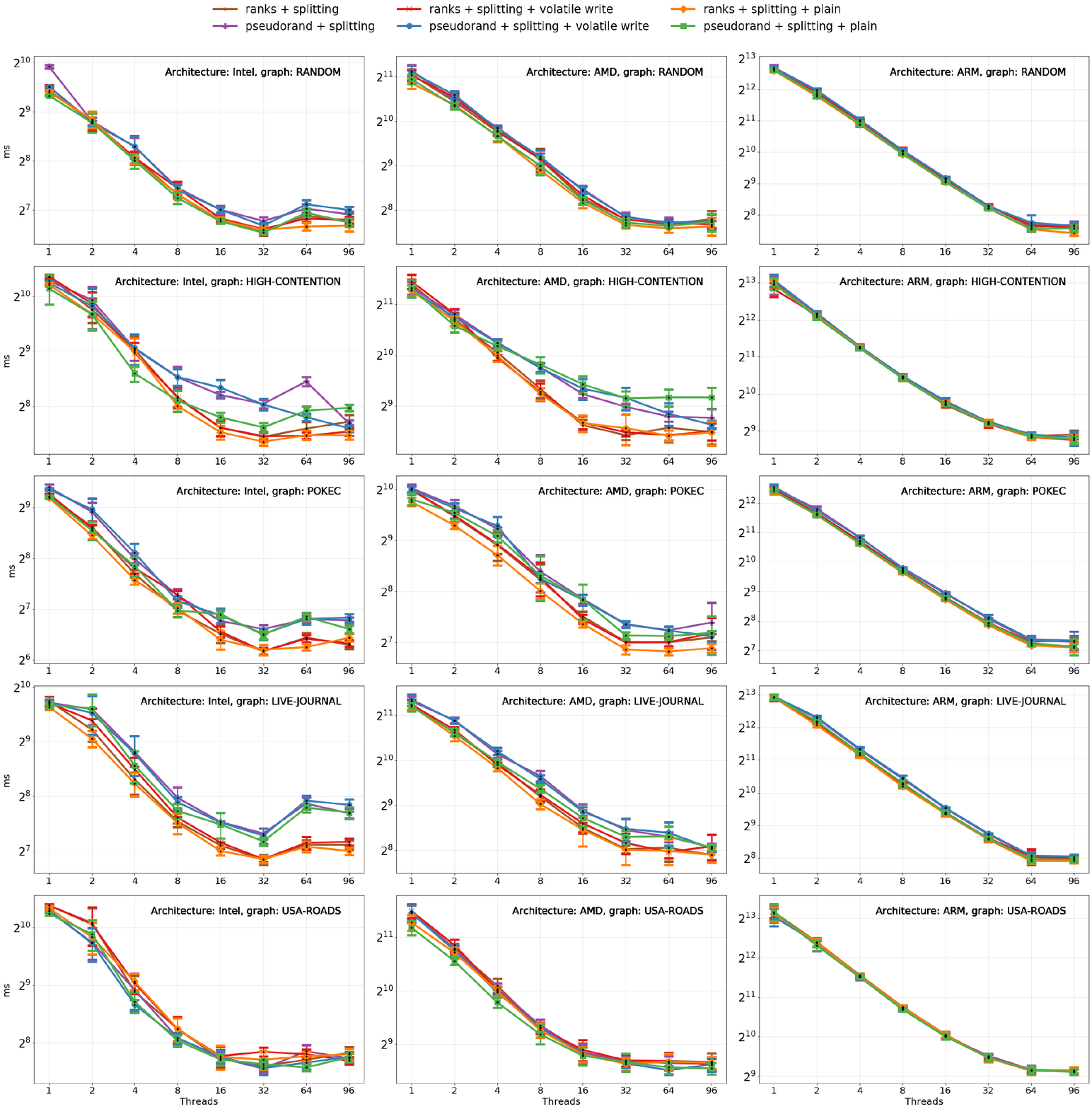}
	\vspace{-9.5em}
\caption{Volatile and plain writes instead of CAS operations to update parents. Here we see, that volatile writes does not perform better that CAS operations, but plain operations give a speedup up to 40\%.}
\label{fig:write-instead-of-cas}
\end{figure}

Since path compaction is a heuristic that influences performance but not correctness, we suggest trying to use simple writes (with or without memory barriers) for updating parent links. What is more, when using writes without memory barriers, we can also use reads without memory barriers at \texttt{Find} operations. While this trick is widely-known, our observation is that it does not break the correctness of the DSU algorithm. The only place where we need the reads with memory barriers is the moment when we check that a node is a root. If the check with the parent obtained from a plain read succeeded, then we should re-check using memory barriers.  Figure~\ref{fig:write-instead-of-cas} shows the comparison of versions with volatile and plain writes instead of CAS. While the version with volatile memory access does not show significant improvement, the one without memory barriers is faster to up to \texttt{40\%}. Despite the fact that this optimization can decrease performance on a large number of threads on synthetic graphs, the best algorithm in these scenarios uses rank-based linking strategy, but with this optimization as well. Thus, we consider it useful in almost all further experiments.

\begin{figure}
	\centering
	\vspace{-5em}
	\includegraphics[width=\textwidth]{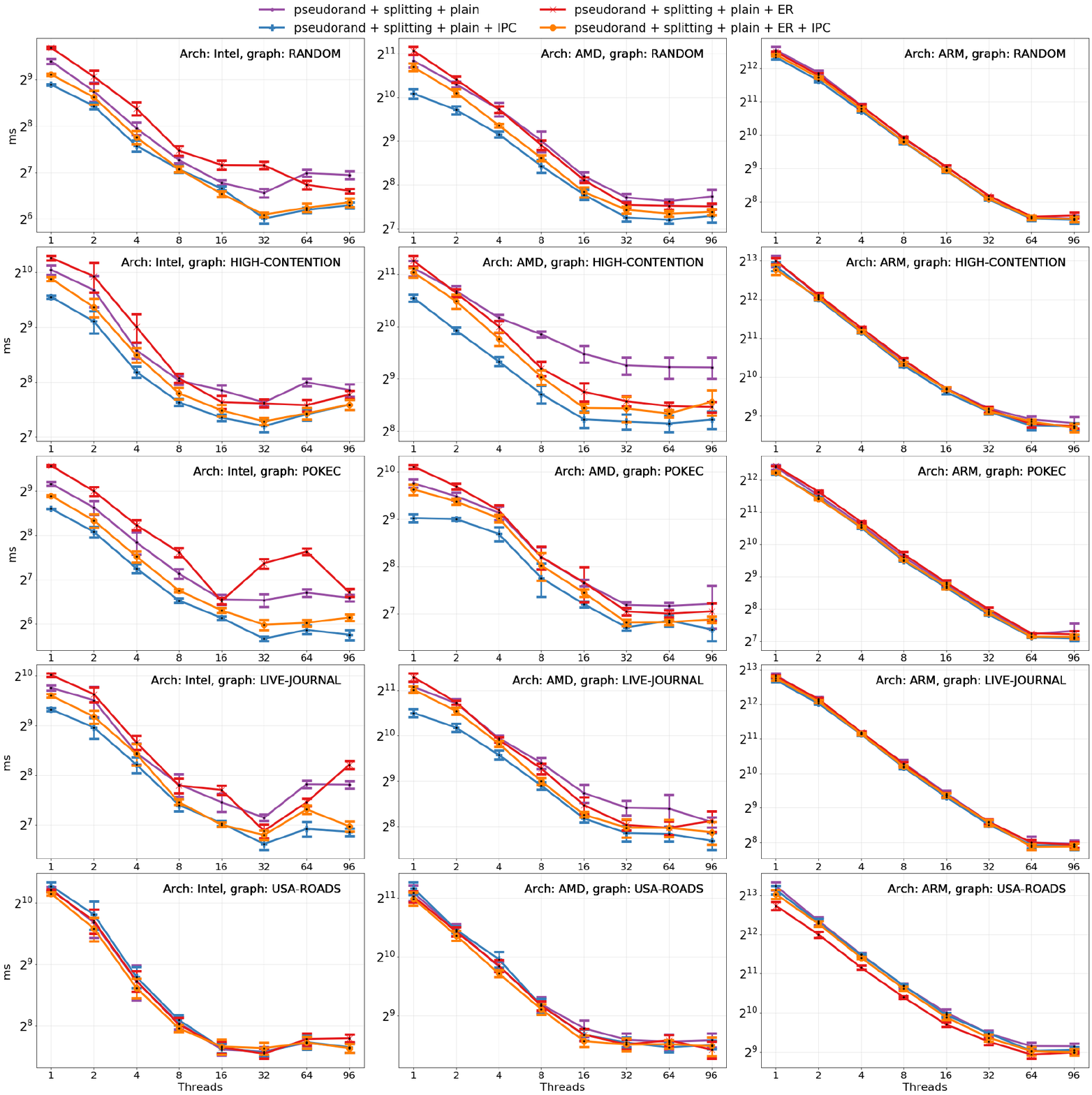}
	\vspace{-9.5em}
\caption{Immediate parent check (IPC) and early recognition (ER) optimizations applied to algorithms with the pseudo-random linking strategy and path compaction via splitting on the connected components benchmark. Here we see that the variant with IPC and without ER wins in all scenarios. We omit the rank-based linking strategy and the Boruvka's algorithm benchmark since the results are similar.}
\label{fig:ipc}
\end{figure}

\paragraph*{Early Recognition.}
The standard \texttt{Union} and \texttt{SameSet} implementations perform two \texttt{Find} operations on their parameters as the first step. However, if both elements are located in the same set, it could be more efficient to terminate the global operation when both \texttt{Find} invocations reach the lowest common ancestor; moreover, it potentially reduces the number of cache misses, which is especially important for NUMA architectures. Similarly, if these two elements are in different sets, it can also be detected during simultaneous climbs {---} it is guaranteed when the first \texttt{Find} reaches the root while the second one stays at an element with greater priority or vice versa. In this situation, \texttt{SameSet} can safely return \texttt{false}, while \texttt{Unite} can link the first root to the current element of the second \texttt{Find} climb. Goel et al. used this technique in sequential versions \cite{Tarjan2014}, while Jayanti and Tarjan adopted it for the concurrent environment \cite{Jayanti2019}. To use the heuristic, we should know priorities for all nodes, so we can not store them only for roots as we have done before for rank-based strategies; thus we will utilize twice more memory.  
Figure~\ref{fig:ipc} shows that this optimization makes the algorithm faster for some scenarios, while the performance becomes worse or the same on others.

\paragraph*{Immediate Parent Check.}
Osipov et al. showed that it is likely for two elements to be united having the same parent links \cite{Osipov2009TheFM} (e.g., if the path is almost fully compressed after lots of \texttt{Find} invocations). Therefore, they suggested using an \emph{immediate parent check (IPC)}  optimization, which examines whether \texttt{parent[u]} and \texttt{parent[v]} are equal in the beginning of \texttt{SameSet} and \texttt{Union} operations and if they are, immediately returns. Figure~\ref{fig:ipc} shows that \emph{immediate parent check} significantly improves the performance of all the considered algorithms, especially on the Intel platform. However, the combination of both early recognition and immediate parent check optimizations work worse due to increasing the code complexity on some graphs and architectures. The immediate parent check optimization is so efficient on our graphs due to having not many connectivity components. The best version presented in the figure is the one that has the \emph{IPC} optimization, but not the \emph{early recognition} one. 

\subsection{Rem's algorithm}

\begin{figure}
	\centering
	    \vspace{-5em}
	\includegraphics[width=\textwidth]{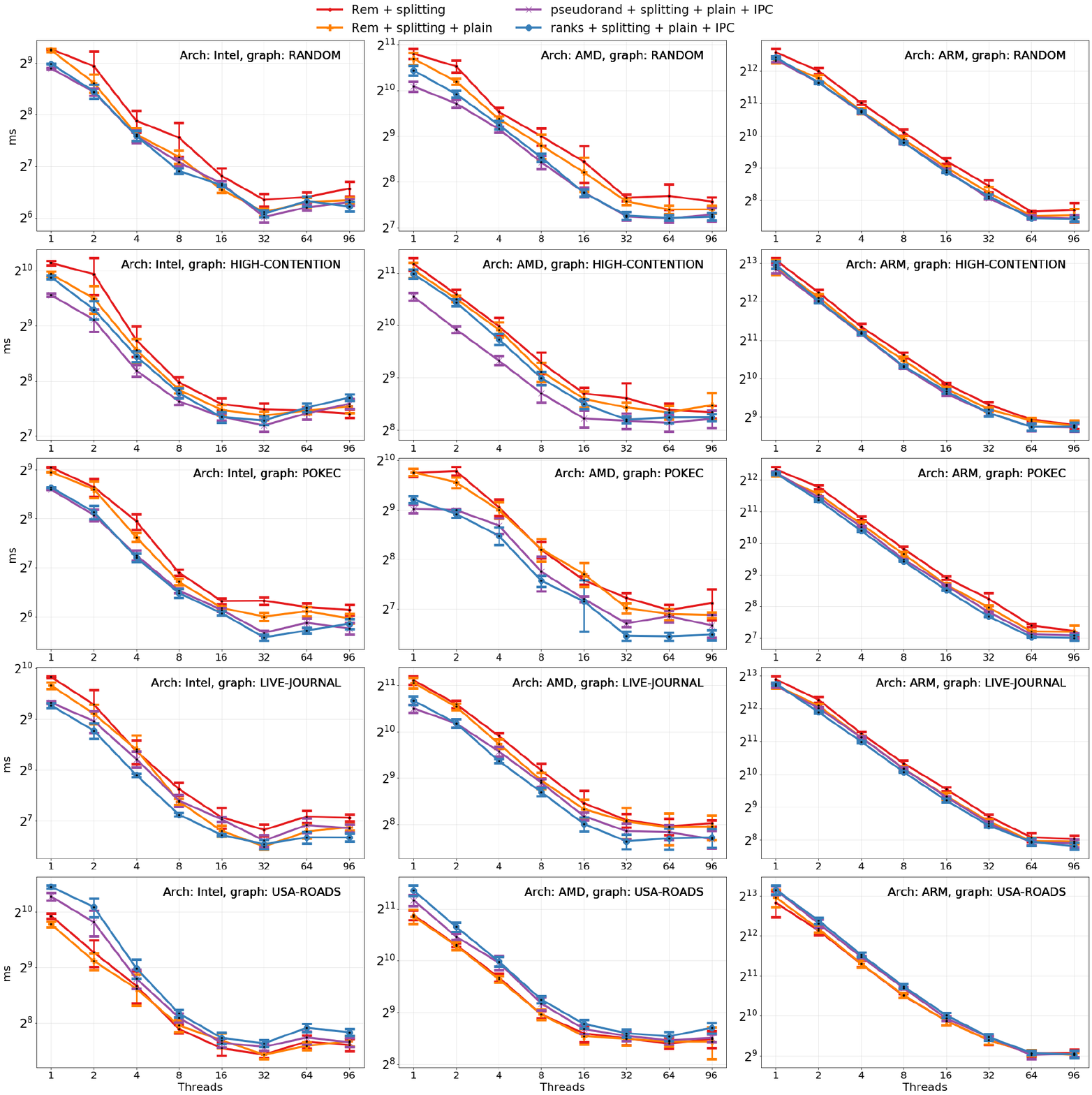}
	\vspace{-9.5em}
\caption{Different variants of Rem's algorithm on the connected components benchmark compared with the previous best algorithms. While Rem's algorithm is the best in the sequential environment, the best of the previously examined ones are slightly better than its concurrent variants.}
\label{fig:rem}
\end{figure}

Rem's algorithm was published in Dijkstra's book ``A discipline of programming'' \cite{Rem}. To our knowledge, it is the first known interleaved algorithm, mixing early recognition and a specific linking strategy. The parents are the priorities of nodes. Then we do not need any memory optimizations, because we need nothing but parents. Similarly as in the early recognition algorithm, we perform two searches of roots, concurrently lifting at each step a node with a lower priority. A concurrent version of Rem's algorithm designed by us is presented at listing~\ref{lst:rem}. Rem's algorithm was identified to perform best by Manne and Patwary in the sequential case~\cite{Patwari}. While it is the best in the sequential environment, Figure~\ref{fig:rem} shows that the previously considered algorithms slightly outperform it. 

\begin{figureAsListing}
\begin{minipage}[t]{0.48\textwidth}
\begin{lstlisting}[mathescape] 
func Unite(u, v) = while (true) {
  up = parent[u]; vp = parent[v]
  if u == v or up == vp: return 
  if vp < up: 
    swap(&u, &v); swap(&vp, &up)
  if u == up:
    if CAS(&u.parent, u, vp):
      return true
  v = parent[up]
  if up != v:
    CAS(&u.parent, up, v)
  u = up
}
\end{lstlisting}
\end{minipage}
\hfill
\begin{minipage}[t]{0.48\textwidth}
\begin{lstlisting}[firstnumber=14,mathescape]
func SameSet(u, v) = while(true) {
  up = parent[u]
  vp = parent[v]
  if u == v or up == vp:
    return true
  if vp < up:
    swap(&u, &v); swap(&up, &vp)
  if u == up: return false
  v = parent[up]
  if up != v:
    CAS(&u.parent, up, v)
  u = up
}
\end{lstlisting}
\end{minipage}
\caption{Concurrent Rem's disjoint-set union algorithm with path splitting}
\label{lst:rem}
\end{figureAsListing}

Rem's algorithms are still better than most of the previously examined algorithms, and there are several reasons for it:
\vspace{-0.5em}
\begin{itemize}
    \item It uses the \emph{early recognition} optimization.
    \item Since we need to compare priorities, we already know the parents at each step and can check them for equality having an automatic \emph{immediate parent check} at each step.
    \item Its priorities are highly correlated with the distance to a root, which is important for \emph{early recognition}. Random priorities never change, ranks change only during \texttt{Union} invocations, while in Rem's algorithms priorities change after each path compaction, when a distance to the root decreases.
\end{itemize}

\subsection{Transactional Memory}
Since the number of failed CAS operations is imperceptible (see Figure~\ref{fig:failed-cas}), it is reasonable to expect that concurrent operations work with different memory locations almost all the time. Therefore, the union-find problem is potentially a good application for hardware transactional memory (HTM); we expect that almost all transactions will succeed. Instead of trying to use HTM as a fast-path for the concurrent versions, we check whether the best sequential algorithms can be scalable using a coarse-grained lock with the lock elision via HTM optimization \cite{rajwar2001speculative}. The TM-based algorithms are coupled with all the sequential algorithms we discussed, and with all the corresponding sequential optimizations.

\begin{figure}
	\centering
	\includegraphics[width=\textwidth]{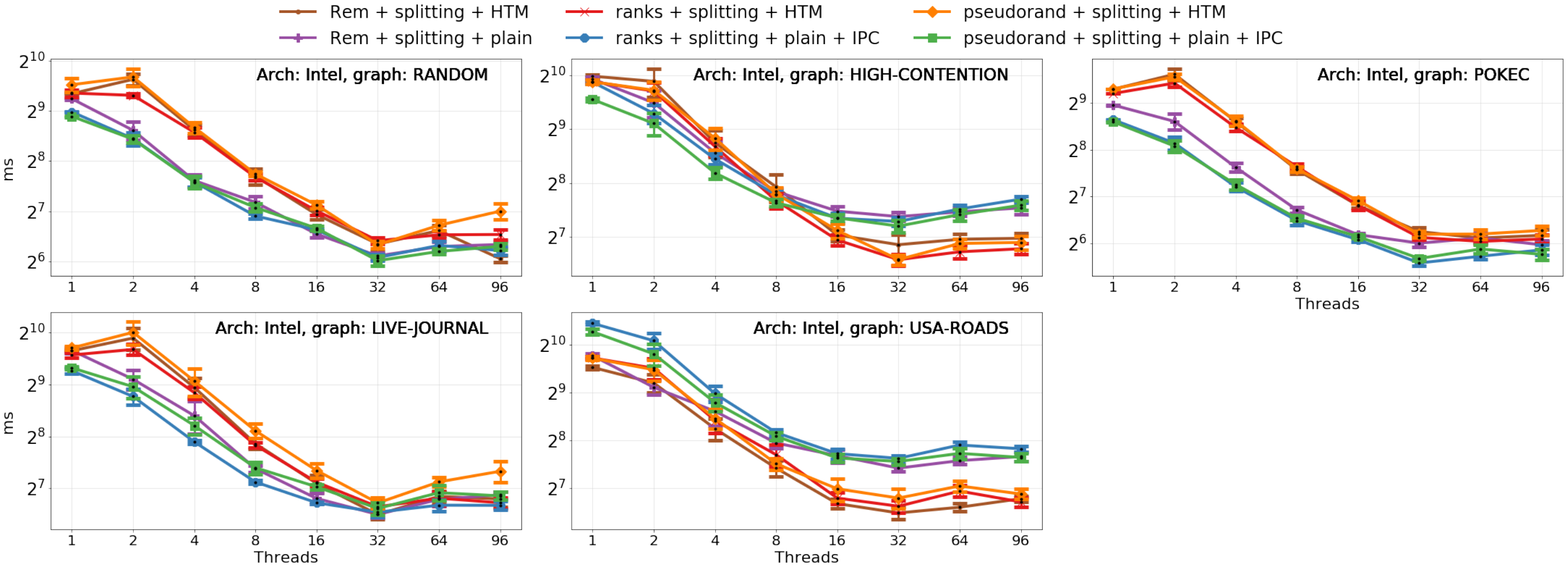}
	\vspace{-7.5em}
\caption{Comparison of lock-based implementations with lock elision via HTM and the best previous algorithms on the connected components benchmark. The versions with HTM are significantly better on two of the considered scenarios and not much worse on other ones, being much simpler at the same time. Graphs, where the results are better, are sparse, so transactions are less likely to touch the same memory.}
\label{fig:htm}
\end{figure}

\begin{figure}
	\centering
	\includegraphics[width=\textwidth]{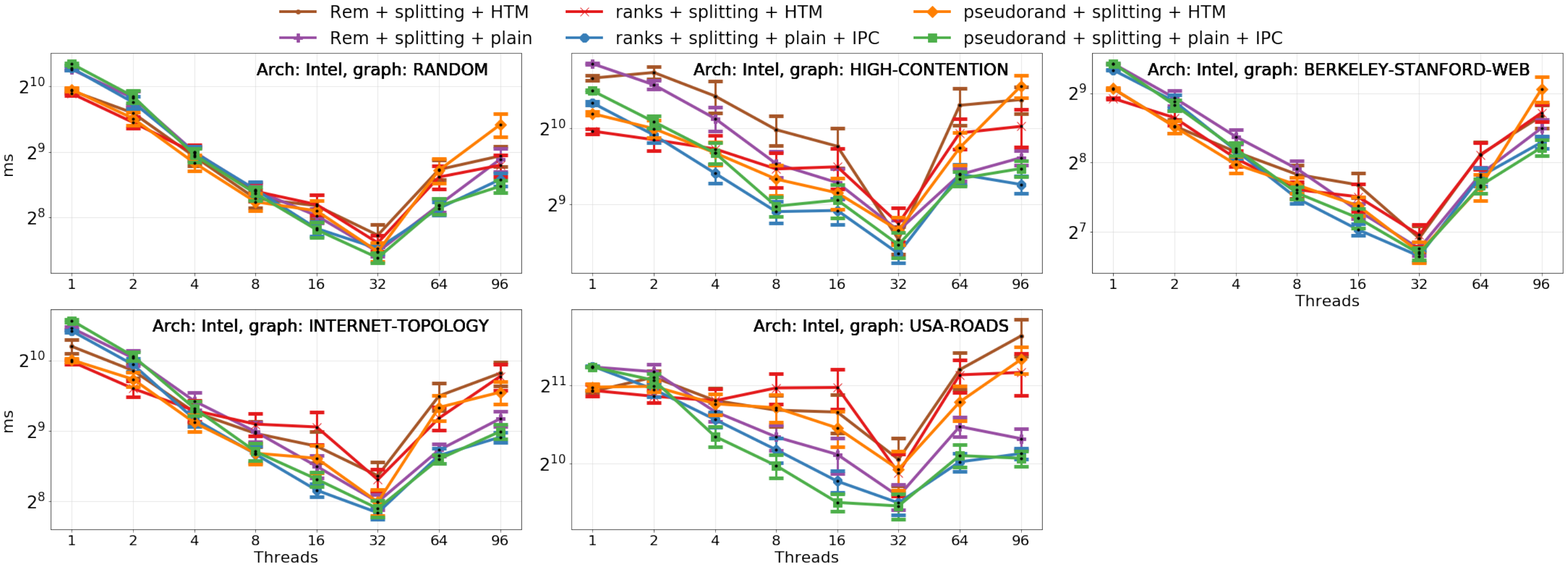}
	\vspace{-7.5em}
\caption{Comparison of lock-based implementations with lock elision via HTM and the best previous algorithms on the Boruvka's algorithm benchmark.}
\label{fig:boruvka-htm}
\end{figure}

Figures~\ref{fig:htm} and~\ref{fig:boruvka-htm} show results evaluated on our local Intel server; unfortunately, AMD and ARM do not support transactional memory at this point. The versions with HTM are definitely superior in some of the considered scenarios, being much simpler at the same time. The reason that for \texttt{USA-ROADS} and for \texttt{HIGH-CONTENTION} graphs on the connected components benchmark they perform better is that these graphs are sparse and operations are unlikely to touch the same memory. Therefore, we conjecture that as general rule algorithms with HTM perform better for sparse graphs. At the same time, the fastest versions in other scenarios use either pseudo-random or rank priorities with plain memory accesses and the immediate parent check optimization; the difference between them is insignificant.

\section{Discussion}
\label{sec:discussion}

We were the first to perform a thorough study of the practical performance of concurrent DSU implementations, exploring architectures (Intel, AMD, ARM), implementation variants (e.g., baseline, Rem), compaction strategies, as well as optimizations and synchronization techniques (lock-free, HTM). We mention the following salient points. 

\paragraph*{Memory Transfer Cost.} 
One basic (and predictable) conclusion of our study is that one of the critical factors for the performance of DSU implementations is the amount of memory totally used and accessed per operation, and, in particular, the average number of cache misses per operation. This is natural since the computational cost of this workload is fairly negligible relative to the cost of the memory traversals. 
Thus, across architectures and algorithmic variants, implementations that try to minimize memory accesses provide better performance. For instance, this is the case of the pseudo-random priorities optimization, which trades off a trivial amount of extra computation for reduced memory transfer. 

Another point is that compression by rank (subtree height) appears to perform particularly well with respect to other methods (in particular, pseudo-random ranks) since it tends to produce higher-quality compressed trees. Notice that this diverges from the theoretical proposals of~\cite{Jayanti2016, Jayanti2019}, which avoid ranks due to additional algorithmic complexity.  

\paragraph*{Synchronization Cost.} 
The second observation regards the synchronization cost of this problem: since most of the overhead of individual operations comes from synchronization cost (e.g., CAS for path compression), it is natural to investigate \emph{relaxing} the strength of the synchronization primitives to improve performance. 
This is the primary motivation behind replacing CAS with plain or volatile write as an optimization, which does appear to lead to noticeable performance improvements. 
We note that, in this case, our empirical observations diverge from the theory; for instance, one of the algorithmic proposals of~\cite{Jayanti2016} requires a repeated CAS sequence to prove the performance upper bound, whereas our results clearly show that the simple update is sufficient in practice. 

\paragraph*{Compaction Quality.}
One of our objectives has been to investigate how the various heuristics for path compaction differ in terms of their practical performance. On this point, we find ourselves unable to adjudicate a clear winner between halving and splitting, which tend to provide similar performance. We do consistently observe that classic path compression leads to worse performance, which is to be expected due to the higher cost. 

\paragraph*{Scalability.} 
In terms of scalability, we note that many graph algorithms are scalable up to the point where NUMA effects come into play, that is, at the point where threads might be accessing memory outside of their socket. This is especially salient in the case of the Boruvka application, where the cost of merging components across sockets dominates in the multi-socket case. 
We leave as an exciting future work direction the question of whether provably efficient concurrent NUMA-aware algorithms exist.

\paragraph*{Performance Across Architectures.} 
We notice that performance trends are consistent across architectures.
Two noticeable differences are the better multi-socket performance of AMD processors (where NUMA effects become detrimental after 3 sockets are used as opposed to just 2 for both Intel and ARM), and the fact that the plain write optimizations work better on Intel and AMD. The last observation befalls because ARM processors are slower, while the memory is as fast as in other machines, and we primarily optimize memory accesses.

\paragraph*{Hardware Transactional Memory.} 
The outstanding performance of HTM solutions was somewhat surprising for us. We explain this in hindsight by the simple fact that HTM allows the algorithm designer in this case to implement the simplest, most fast-path-efficient variant of the algorithm (in particular, Rem), which avoids most of the average synchronization costs of other techniques and allows more compiler optimizations, while contention is fairly limited in most scenarios. 

\paragraph*{Performance versus Monetary Cost.} 
Since we ran our benchmarks on cloud instances, we are able to provide some intuition also with respect to the average cost per experimental run. In this case, the clear winner is the ARM machine, which offers significantly cheaper CPU time, at a very high core count.  

\paragraph*{The Fastest Concurrent Union-Find Algorithm.} 
We conclude that, within the confines of our experimental setup, our optimized concurrent variants with either splitting or halving path compaction strategy and pseudo-random or rank priorities are best across architectures and inputs with an insignificant difference. 
This finding is perhaps disappointing from the theoretical perspective since we use plain writes instead of several attempts to update parents. Considering that we also use the immediate parent check optimization, it would be interesting to further study the guarantees of the resulting algorithms in the context of concurrent executions, perhaps for common distributions of (non-adversarial) input graphs. 

\bibliographystyle{unsrt} 
\bibliography{references}

\appendix

\end{document}